\documentclass[twocolumn,preprintnumbers,amsmath,amssymb,graphicx,prb]{revtex4}

\usepackage{graphicx}

\begin{document}

\title{Novel growth mechanism of epitaxial graphene on metals}

\author{Andrew Zangwill}
\affiliation{School of Physics, Georgia, Institute of Technology, Atlanta, GA 30332}

\author{Dimitri D. Vvedensky}
\email{d.vvedensky@imperial.ac.uk}
\affiliation{The Blackett Laboratory, Imperial College London, London SW7 2AZ, United Kingdom}

\date{\today}

\maketitle

Graphene, a hexagonal sheet of $sp^2$-bonded carbon atoms, has extraordinary properties which hold immense promise for future nanoelectronic applications.  Unfortunately, the popular  preparation methods of  micromechanical cleavage and chemical exfoliation of graphite do not easily scale up for application purposes.  Epitaxial graphene provides an attractive  alternative, though there are many challenges, not least of which is the absence of an understanding of the complex atomistic assembly kinetics of graphene. Here, we present a simple rate theory of epitaxial graphene growth on close-packed metal surfaces. Based on recent low-energy electron-diffraction microscopy experiments (LEEM) \cite{loginova09}, our theory supposes that graphene islands grow predominantly by the addition of five-atom clusters, rather than solely by the capture of diffusing carbon atoms.   With suitably chosen kinetic parameters, our theory produces a time-dependent carbon adatom density that is in quantitative agreement with measured data. The temperature-dependence of this adatom density at the onset of nucleation leads us to predict that the smallest stable precursor to graphene growth is an immobile island composed of six five-atom clusters.  Our findings provide a starting point for more detailed simulations which will yield important input to developing strategies for the large-scale production of epitaxial graphene.

The epitaxial growth of graphene on transition-metal surfaces has been studied by surface scientists for many years  \cite{gall97}. The broader scientific community took notice of graphene in 2004, when two groups independently pointed out its unusual transport properties and potential for microelectronic applications \cite{novoselov04,berger04}.  Subsequent work has revealed additional remarkable properties, such as anomalies in the integer quantum Hall effect \cite{novoselev05} and in quasi-particle coupling \cite{bostwick07}, which are signatures of charge carriers that behave as massless Dirac fermions.

The exfoliation technique  of Ref.~[\onlinecite{novoselov04}] is often used to produce samples of graphene for two-dimensional electron gas studies.  This method  is quick and easy, but it is uncontrolled and unsuitable for scale-up to the production volumes that will be necessary for the most interesting and important commercial applications. An attractive alternative is to grow graphene epitaxially on a hexagonal substrate. Considerable progress has been achieved in this direction, both using silicon carbide \cite{first10,sprinkle10} and close-packed metals \cite{wintterlin09} as the substrate. Nevertheless, almost nothing is known about the atomistic formation kinetics of epitaxial graphene on any substrate.  

The present paper is motivated by particularly elegant LEEM experiments reported by Loginova and co-workers for the growth kinetics of the Gr/Ru(0001) system \cite{loginova08}.  Using electron reflectivity to monitor the evolution of the density of carbon adatoms deposited either from a heated carbon rod or from an ethylene source, it was discovered  that the step edges of two-dimensional (2D) graphene islands advance with a velocity $v$ that varies as the {\it fifth} power of the ratio of the carbon adatom density $n$ to its equilibrium value $n_{\rm eq}$:
\begin{equation}
v=k\biggl[\biggl({n\over n_{\rm eq}}\biggr)^5-1\biggr]\, .
\label{eq1}
\end{equation}
This led the authors to suggest that the islands grow by the attachment of five-atom clusters, rather than by the usual mechanism of single adatom attachment.  More recent work by the same group extends this conclusion to the Gr/Ir(111) system \cite{loginova09}.

In this paper, we develop a rate theory for the epitaxial growth of graphene on metal surfaces.  Rate equations have been used for some time for modelling epitaxial systems \cite{venables84} and for establishing general principles such as the existence of scaling regimes for concentrations of surface species and scaling forms for island-size distributions \cite{evans06}.  Scanning tunneling microscopy images, both as ``snapshots'' of quenched surfaces and {\it in situ} scans \cite{voigtlander01}, are usually interpreted in terms of kinetic Monte Carlo simulations because of the wealth of spatial information contained in such images \cite{kotrla02}.  Nevertheless, rate equations continue to  provide the conceptual framework for interpreting essentially all growth scenarios.   The amount and quality of time-resolved data in the present case is unprecedented and, as we will show, the  temperature dependence of the carbon adatom density at the onset of nucleation and at equilibrium encode key information about the atomistic processes of graphene formation on Ru(0001) \cite{loginova09} that would be unavailable without a theoretical analysis.

A simple model based on the experiments of Loginova {\em et al.}~\cite{loginova08} assumes that carbon atoms arrive at the surface with a flux $F$ and migrate across the surface with diffusion constant $D$.  Adatoms can attach to the perimeter of 2D islands, but this is not the predominant mechanism of island growth.  Instead, clusters composed of $i=5$ carbon atoms form when $i$ adatoms collide.  These clusters migrate with surface diffusion constant $D^\prime$ and, when $j=6$ such clusters collide, an island of size $i\times j$ is formed.   Islands grow mainly by the attachment of $i$-atom clusters.  The value $i=5$ for the size of mobile clusters is derived from experiment, as noted above, while the value $j=6$ is obtained from a fit to the temperature dependence of the carbon adatom density at the onset of nucleation.  The latter is a new and striking result, and suggests that the collision of fewer than six clusters leads to the formation of transient, possibly mobile, species whose structure bears little resemblance to graphene.  Even islands formed from six mobile clusters are unlikely to immediately transform into graphene, but instead pass through a series of intermediate structures. This general scenario is consistent with tight-binding, grand canonical Monte Carlo simulations which produce small carbon chains as a precursor to graphene formation on Ni(111) \cite{amara06}.

The rate equations for these kinetic steps are expressed in terms of the homogeneous densities of
carbon adatoms $n$, five-atom clusters $c$, and islands $N$ as
\begin{align}
\label{eq2}
{dn\over dt}&=F-iDn^i+iKc-DnN+K^\prime N\, ,\\
\noalign{\vskip3pt}
{dc\over dt}&=Dn^i-Kc-D^\prime cN-jD^\prime c^{\,j}\, ,\\
\noalign{\vskip3pt}
{dN\over dt}&=D^\prime c^{\, j}.
\label{eq4}
\end{align}
In these equations, $K$ is the cluster dissolution rate and $K^\prime$ is the detachment rate of adatoms from islands. For simplicity, we have omitted refinements to these equations, such as capture numbers, that make them more realistic \cite{evans06,ratsch03}, but also more numerically cumbersome \cite{amar01}.  Thus, our proposed rate equations are  sufficiently minimal to enable a straightforward optimization of parameters, while retaining enough physical content to permit a detailed interrogation of experimental data.  We will demonstrate below that this set of equations is capable of providing a quantitative account of all of the measured data reported in
Ref.~[\onlinecite{loginova09}].  Moreover, our analysis will reveal the presence of new kinetic mechanisms for graphene formation on metal surfaces.

Each of the rates parameters $D, D^\prime, K$, and $K^\prime$ is taken to have the Arrhenius form $\nu_0e^{-\beta E}$ with the prefactor $\nu_0=2k_BT/h$ assumed to be the same for all processes. Otherwise, $k_B$ is Boltzmann's constant, $T$ is the absolute temperature of the substrate, $h$ is Planck's constant, $\beta=1/(k_BT)$, and $E$ is the energy barrier to the process.  For the temperature range of interest (790~K--1050~K), $\nu_0\sim10^{13}$~s$^{-1}$.  We have set the lattice constant $a=1$, but the required factors of $a$ can be reinstated into (\ref{eq2})--(\ref{eq4}) by straightforward dimensional analysis.

\begin{figure*}
\centering
\includegraphics[width=5.75cm]{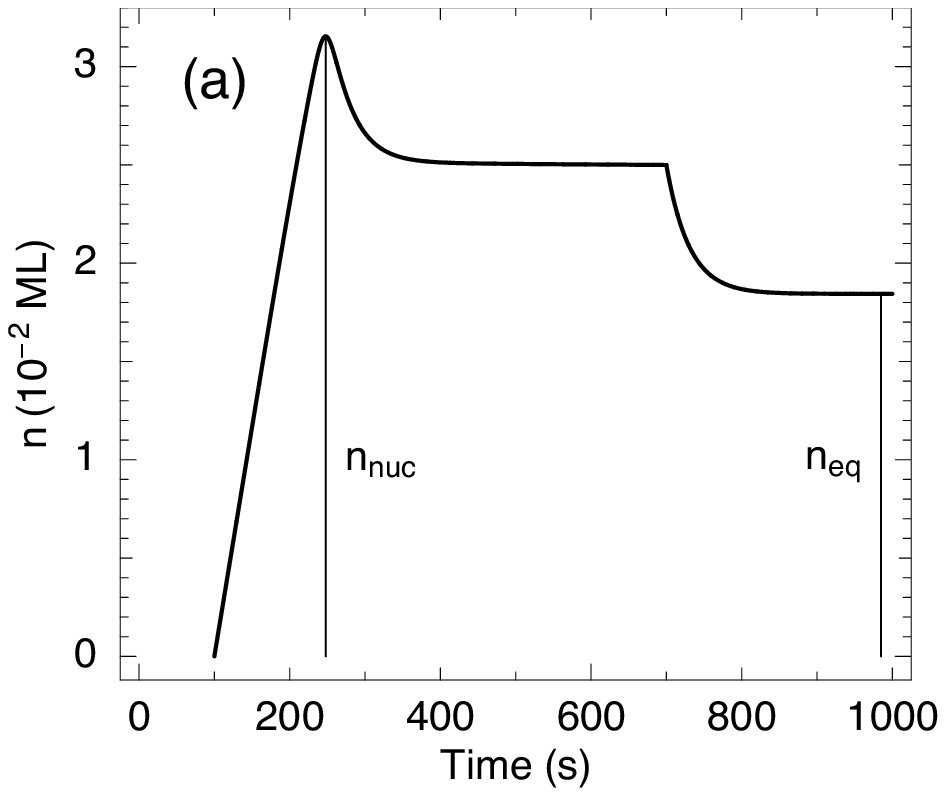}\hskip0.25cm
\includegraphics[width=5.75cm]{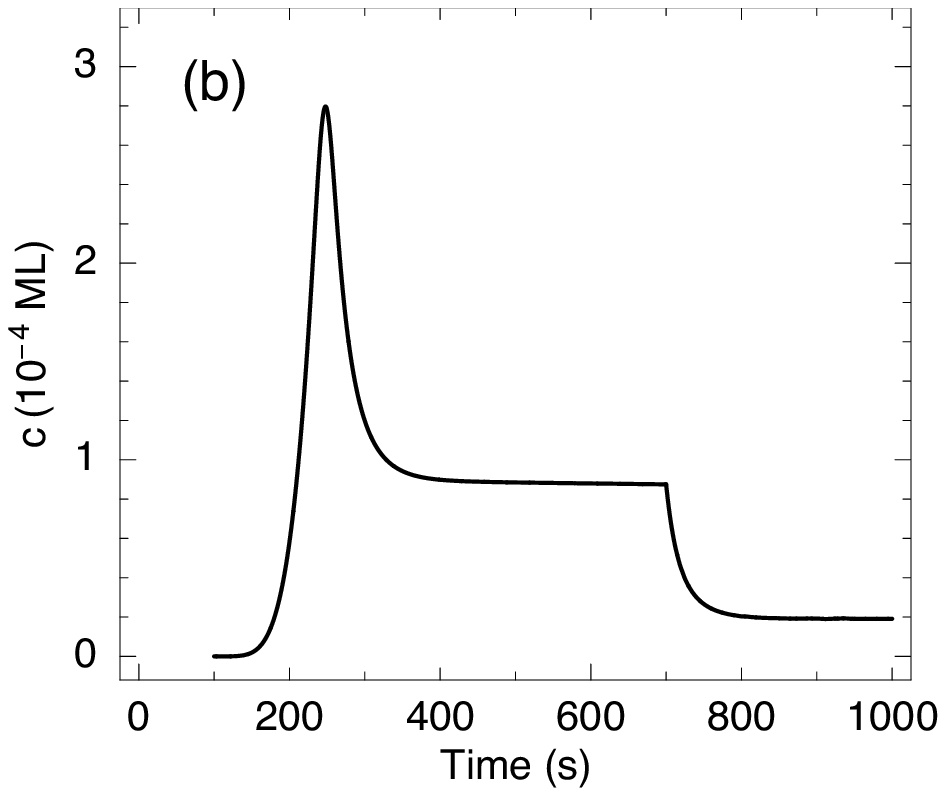}\hskip0.25cm
\includegraphics[width=5.75cm]{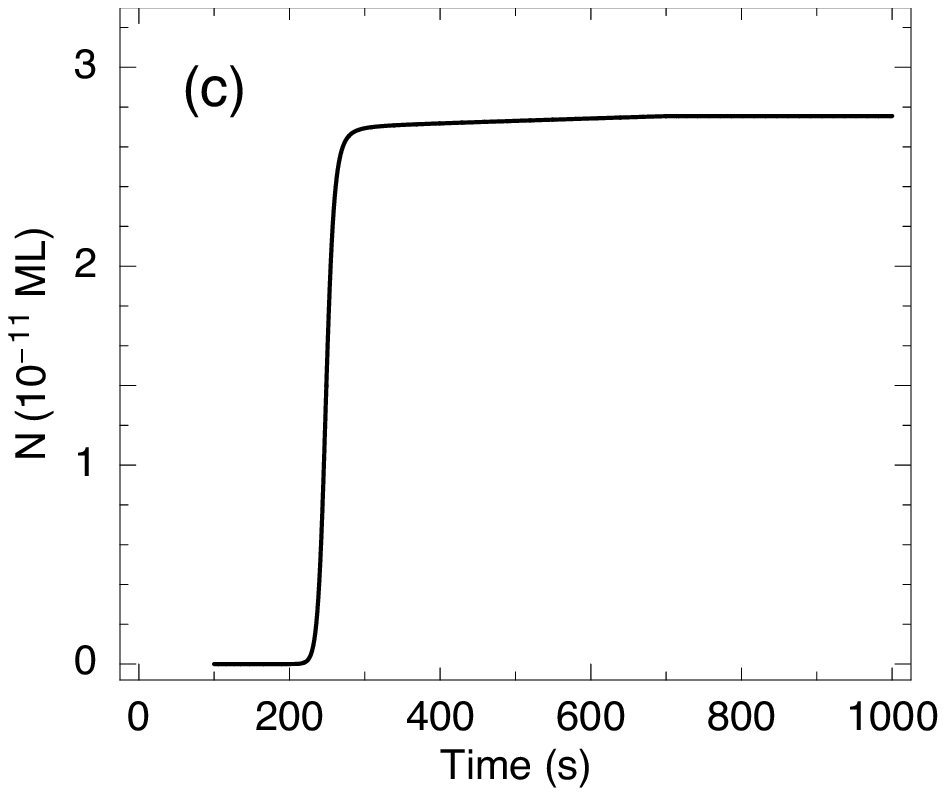}
\caption{{\bf Solution of rate equations.} Coverage dependence of (a) the carbon adatom density $n$, (b) the five-atom mobile cluster density $c$, and (c) the total immobile island density $N$ obtained from the rate equations (\ref{eq2})--(\ref{eq4}) at $T=1020$~K.  The adatom densities at the onset of nucleation $n_{\rm nuc}$ and at equilibrium $n_{\rm eq}$ are indicated in (a).  The kinetic parameters are $E_D=0.92$~eV, $E_{D^\prime}=0.87$~eV, $E_K=1.72$~eV, $E_{K^\prime}=1.27$~eV, and $j=6$.}
\label{fig1}
\end{figure*}

There are four energy barriers that must be determined, as well as the value of $j$.  However, this is not an unconstrained optimization, as there are restrictions which limit the ranges of these parameters.  For example, the equilibrium carbon adatom concentration $n_{\rm eq}$ [Fig.~\ref{fig1}(a)], determined from the time-independent solution to (\ref{eq2}) with $F=0$,  yields
\begin{equation}
n_{\rm eq}\sim {K^\prime\over D}=e^{-\beta(E_{K^\prime}-E_D)}\, .
\label{eq6}
\end{equation}
As $n_{\rm eq}$ is found to be an increasing function of temperature \cite{loginova09}, we must choose $E_{K^\prime}>E_D$.   Indeed, by fitting this form of $n_{\rm eq}$ to the corresponding experimental values, we find that $E_{K^\prime}-E_D=0.35$~eV.  In obtaining (\ref{eq6}), we have used the fact that $n\gg c\gg N$, which results from the unusually large values of $i$ and $j$.  This can be checked afterward for self-consistency (Fig.~\ref{fig1}).

Further insight into the growth parameters is obtained by examining the stationary concentration $n_s$ in the presence of a non-zero flux.  Again invoking the ordering $n\gg c\gg N$, we find,
\begin{equation}
n_s^{ij+1}=(ij+1)^{-1}\biggl({DD^\prime\over F^2}\biggr)^{-1}\biggl({D\over K}\biggr)^{-j}\theta^{-1}\, ,
\end{equation}
where $\theta=Ft$ is the coverage. At fixed $\theta$, the tem\-perature-dependence of $n_s$ is
\begin{equation}
n_s\sim \exp\biggl\{{\beta\over ij+1}\Bigl[E_{D^\prime}-jE_K+(j+1)E_D\Bigr]\biggr\}\, .
\label{eq7}
\end{equation}
As the behavior with temperature of $n_s$ is qualitatively similar to that of $n_{\rm nuc}$, which is observed to be an {\it increasing} function of temperature \cite{loginova09}, we must require that
\begin{equation}
E_{D^\prime}-jE_K+(j+1)E_D<0\, .
\label{eq11}
\end{equation}
Note that this condition involves $j$ but not $i$.  For our energy parameters, we must choose $j>2$.  Beginning with five parameters ($E_D$, $E_{D^\prime}$, $E_K$, $E_{K^\prime}$, and $j$), the restrictions in (\ref{eq6}) and (\ref{eq11}) reduce our optimization to an effective three-parameter fit.

The solutions of the rate equations for the evolutions of the densities of the carbon adatoms, the five-atom clusters, and the immobile islands are shown in Fig.~\ref{fig1} with the optimized parameters in Table~\ref{table1}.  These solutions are most sensitive to the energy barriers for $D$, $K$, and $K^\prime$, with variations of $\pm0.01$~eV producing noticeable changes in the solutions.  In contrast, variations by as much as $\pm0.05$~eV of the diffusion barrier $D^\prime$ for the five-atom clusters produce minimal changes, especially for the adatom solution.  This is due, in part, to the fact that the processes that most influence the adatom density are the formation and dissolution of the clusters.  The density of islands and their sizes have a smaller effect because the attachment and detachment processes between adatoms and islands occur much less frequently, as the island step velocity in (\ref{eq1}) indicates.

Most apparent from the solutions in Fig.~\ref{fig1} is how qualitatively different the adatom and island profiles are from those found in the ``usual'' epitaxial growth scenario where  only adatoms are mobile and there is a critical size for island formation \cite{evans06,bales94}.  The presence of mobile clusters  composed of five atoms and the large size $(j=6)$ of the smallest immobile islands formed from  five-atom clusters are  responsible for this behavior.  The clusters and islands both show sudden increases in their densities due to bursts of aggregation when there is a sufficient adatom density to first form five-atom clusters.   The onset of island formation is further delayed until there are enough five-atom clusters so that a six-cluster collision becomes likely.  The islands are formed over a short time period after which the island density sasturates with essentially no additional nucleation.

\begin{figure*}
\centering
\includegraphics[width=7cm]{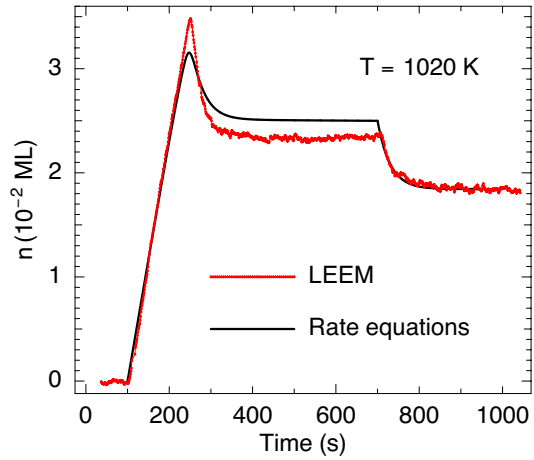}
\caption{{\bf Rate theory and LEEM measurements for time-dependent carbon adatom coverage.} The evolution of the carbon adatom density in Fig.~\ref{fig1}(a) (black curve) compared with the corresponding LEEM data in Ref.~[\onlinecite{loginova09}] (red points)}.
\label{fig2}
\end{figure*}

Figure~\ref{fig2} shows the solution for the carbon adatom density in Fig.~\ref{fig1}(a) overlayed with the experimental data in Fig.~12 of Ref.~[\onlinecite{loginova09}].  The theoretical solution reproduces all of the main features of the experimental data, though there are small differences in some of the details.  The rate equations account for the position of the nucleation peak at approximately 150~s after the flux is introduced into the system, but underestimate the experimental value of $n_{\rm nucl}$ by $\simeq10\%$.  The subsequent stationary regime is also accounted for by our theory, but the solution overestimates the steady experimental value by a few per cent.  After the cessation of the flux, the theoretical and experimental equilibration curves are indistinguishable.

The most likely cause of the discrepancies between the theoretical and experimental carbon adatom evolutions is the omission of any spatial information in the rate equations.  By using spatially averaged densities for all surface species, the local spatial arrangements of five-atom clusters and islands, which affects adatom attachment and detachment through the notion of ``capture zones'' \cite{evans06,bales94}, and any preferential sites for cluster or island formation, are excluded.  These would first be apparent at the onset of nucleation, where the spatial arrangement of islands is critical to establishing their growth rates \cite{ratsch00}.  The subsequent discrepancy in steady-state regime is then a consequence of this underestimate of the nucleation rate.

On the other hand, the rate equations provide an excellent account of the equilibration of the adatom concentration after the cessation of the incident mass flux.  The comparative immunity of the equilibration regime to spatial correlations between islands suggests that $n_{\rm eq}$ is dominated by the detachment of adatoms from an island followed by re-attachment to the same island.   An immediate consequence of this is that  the distribution of island sizes would not change over the time scale of equilibration.  This can be tested experimentally.

\begin{figure*}
\centering
\includegraphics[width=7cm]{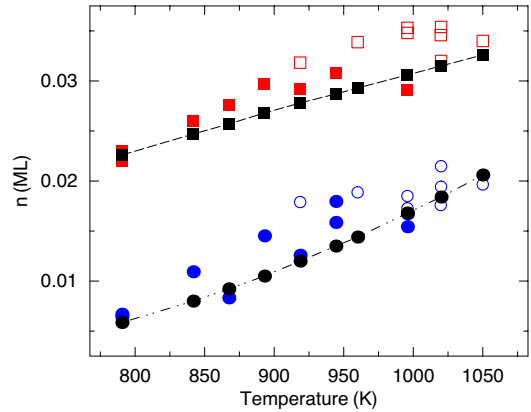}
\caption{{\bf Rate theory and LEEM measurements for adatom coverage at onset of nucleation and at equilibrium.} The temperature dependence of the concentration of carbon adatoms $n_{\rm nuc}$ required to
nucleate graphene (red symbols) and the carbon adatom concentration $n_{\rm eq}$ at equilibrium
(blue symbols) compared with the corresponding quantities obtained from rate equations (black
symbols) as indicated in Fig.~\ref{fig1}.  Experiments \cite{loginova09} were carried out with carbon
vapor (filled red and blue symbols) and ethylene (open red and blue symbols).}
\label{fig3}
\end{figure*}

Figure~\ref{fig3} compares the temperature-dependence of $n_{\rm nuc}$ and $n_{\rm eq}$
[Fig.~\ref{fig1}(a)] obtained from the rate equations with those observed in experiments
\cite{loginova09}.  As Fig.~\ref{fig2} indicates, $n_{\rm eq}$ is expected to be more accurate than $n_{\rm nuc}$, although both quantities provide good accounts of the experimental data.  The computed equilibrium concentrations from the full solution are
indistinguishable from that in (\ref{eq6}), which shows that, while single adatom attachment and detachment do not have a
significant direct impact on island kinetics during growth, they do have experimentally observable consequences during equilibration.   On the other hand, the temperature-dependence of $n_{\rm nuc}$ embodies more information about kinetics during growth, as Eq.~(\ref{eq7}) indicates, especially with regard to the number of five-atom clusters required to form an incipient graphene island.  The choice $j=6$ yields the best fit to the data in Fig.~\ref{fig3}.  This comparison, more than any other reported here, indicates the value of our rate equation analysis.  An increasing $n_{\rm nuc}$ with temperature is not at all typical of epitaxial systems, and its consequences would have been difficult, if not impossible, to ascertain without a theoretical model that abstracts all but the most essential kinetic steps.
\medskip

\noindent
{\bf Methods} 
\noindent
Rate equations are typically formulated phenomenologically for a particular system with basic guidance from experiments and fundamental calculations.  In this case, we began with processes involving the diffusing carbon adatoms and five-atom clusters. Additional processes and all parameters were identified through comparison with the available experimental data.  The mathematical structure of rate equations for epitaxial systems is a system of coupled autonomous nonlinear ordinary differential equations that often  have a property known as ``stiffness,'' which arises from the disparity of rates in the equations.  This means that their solutions become unstable in certain parameter regimes.  Our calculations were carried out with {\sc Mathematica}\cite{wolfram}, which has a solver that can accommodate such cases.

\medskip

\noindent
{\bf Acknowledgements.}  The authors thank Elena (Loginova) Starodub for providing the data in Figs.~\ref{fig2} and \ref{fig3} and both she and Norman Bartelt for an informative correspondence.
\medskip

\noindent
{\bf Author Contributions.} A.Z. conceived this study. The rate equations were formulated jointly by both authors.  The numerical calculations were carried out by D.D.V. and both authors wrote the manuscript.
\medskip

\noindent
{\bf Competing Financial Interests.} The authors declare no competing interests.

\newpage

\newpage

\newpage

\newpage

\begin{table*}
\caption{\label{table1} The optimized energy barriers (in eV) in the Arrhenius forms of $D$, $D^\prime$, $K$, and $K^\prime$ in (\ref{eq2})-(\ref{eq4}).  The accuracy of each barrier refers to the sensitivity of the solutions to variations of that barrier.}
\begin{ruledtabular}
\begin{tabular}{lcc} \vspace{0.05cm}
Rate & Barrier & Accuracy\\ \hline \\[-0.25cm]
$D$ & 0.92 & 0.01\\[0.1cm]
$D^\prime$ & 0.87 & 0.05\\[0.1cm]
$K$ & 1.72 & 0.01\\[0.1cm]
$K^\prime$ & 1.27 & 0.01\\[0.1cm]
\end{tabular}
\end{ruledtabular}
\end{table*}

\end{document}